\documentclass[twocolumn]{aa}

\usepackage{graphicx}
\usepackage{txfonts}
\usepackage{cleveref}   
\usepackage{hyperref}
\makeatletter

\newcommand{\Rmnum}[1]{\expandafter\@slowromancap\romannumeral #1@}
\makeatother
\begin{document}

\title{Possible $\sim$~1 hour quasi-periodic oscillation in narrow-line \\ 
Seyfert 1 galaxy MCG--06--30--15}

\author{
Alok C. Gupta \inst{1,2}
\and Ashutosh Tripathi \inst{3}
\and Paul J.\ Wiita \inst{4}
\and Minfeng Gu \inst{1}
\and Cosimo Bambi\inst{3,5}
\and Luis C. Ho \inst{6,7}
}
          
\institute{ 
Key Laboratory for Research in Galaxies and Cosmology, Shanghai Astronomical Observatory, 
Chinese Academy of Sciences, 80 Nandan Road, Shanghai 200030, China \\
\email{acgupta30@gmail.com}
\and
Aryabhatta Research Institute of Observational Sciences (ARIES), Manora Peak, Nainital $-$ 263 001, India
\and 
Center for Field Theory and Particle Physics and Department of Physics, Fudan University, 220 Handan 
Road, Shanghai 200433, China \\
\email{ashutosh31tripathi@gmail.com}
\and
Department of Physics, The College of New Jersey, P.O.\ Box 7718, Ewing, NJ 08628-0718, USA \\
\email{wiitap@tcnj.edu}
\and
Theoretical Astrophysics, Eberhard-Karls-Universit{\"a}t T{\"u}bingen, D-72076 T{\"u}bingen, Germany
\and 
Kavli Institute for Astronomy and Astrophysics, Peking University, Yi He Yuan Lu 5, Hai Dian District, 
Beijing 100871, China
\and
Department of Astronomy, Peking University, Yi He Yuan Lu 5, Hai Dian District, Beijing 100871, China
           }
          
\date{Received 13 June 2018 / Accepted 19 July 2018}

\abstract{

We found a possible  $\sim$ 1 hour quasi-periodic oscillation (QPO) in a $\sim$ 55 ks X-ray observation 
of the narrow-line Seyfert 1 galaxy MCG--06--30--15 made with the  {\it XMM-Newton} EPIC/pn detector in 
the energy range 0.3 -- 10 keV. We identify a total modulation of $\sim$ 16\% in the light curve and find a 
$\simeq$ 3670~s quasi-period using Lomb-Scargle periodogram (LSP) and weighted wavelet Z-transform (WWZ) 
techniques. Our analyses of eight light curves of MCG--06--30--15, indicated the possible presence of 
an oscillation during one of them. The LSP indicates a statistically significant ($\simeq$ 3$\sigma$) 
QPO detection. A WWZ analysis shows that the signal at this possible roughly 3670~s period is present, 
and rather persistent, throughout the observation; however, a signal around 8735~s is more persistent. 
We briefly discuss models that can produce X-ray QPOs with such periods in narrow line Seyfert 1 galaxies, 
as both other claimed QPO detections in this class of AGN had very similar periods.

}


\maketitle

\section{Introduction}

Detections of quasi-periodic oscillations (QPOs) are very rare in active galactic nuclei (AGN),
but are fairly common in both black hole (BH) and neutron star binaries in the Milky way and nearby
galaxies (Remillard \& McClintock 2006). Over the last decade there have been several 
claims of QPO detections on diverse timescales ranging from a few tens of minutes to hours to days and even 
years, using $\gamma-$ray, X-ray, optical and radio monitoring data of various classes of AGN (Gierli{\'n}ski 
et al.\ 2008; Espaillat et al.\ 2008; Gupta et al.\ 2009; Lachowicz et al.\ 2009; Lin et al.\ 2013; King et al. 
2013; Fan et al. 2014; Sandrinelli et al.\ 2014, 2016a, 2016b; Graham et al.\ 2015; Ackermann et al.\ 2015; 
Pan et al.\ 2016; Bhatta et al.\ 2016;  Bhatta 2017; Li et al. 2017; Xiong et al. 2017; Zhang et al. 2017a, 2017b, 
2018; Hong et al. 2018; and references therein). Since any such QPOs are almost certainly transient, verification 
is difficult and claims are strengthened if more than one technique provides consistent results.

The first significant  QPO  reported for a narrow-line Seyfert 1 (NLSy1) galaxy was in a light curve of 
RE J1034$+$396 (Gierli{\'n}ski et al.\ 2008); this had a roughly 1 hour timescale (3730 s) in the X-ray 
band and was found in {\it XMM-Newton} data. The second QPO detection in a NLSy1 galaxy was more recently  
made by Pan et al.\ (2016) in 1H 0707$-$495; it  again had a period of $\sim$ 1 hour (3800 s) and was 
seen in a {\it XMM-Newton} observation.  We report  a third probable QPO detection 
in a NLSy1 for MCG--06--30--15 ($z = 0.00775$), once again using {\it XMM-Newton} data. MCG--06--30--15 
is well known,  particularly for its broad iron K$\alpha$ line; this line that provides strong evidence 
for the presence of a supermassive black hole (SMBH; e.g.,\ Tanaka et al.\ 1995), which is probably spinning 
rapidly (e.g.,\ Iwasawa et al.\ 1996). Because its H$\beta$ line width (FWHM: full width at half maxima) 
of $1933\pm82$ km s$^{-1}$ is $< 2000$ km s$^{-1}$  (Hu et al.\ 2016), this SMBH is classified as a NLSy1. 
Interestingly, the possible QPO we report on in this work also shows a period of $\sim$ 1 hour. Convincing 
detection and careful characterizations of additional QPOs in these and other NLSy1 galaxies may shed new 
light on the physical processes occurring in these sources and can yield information on their SMBH masses 
and spins (e.g.,\ Abramowicz \& Klu{\'z}niak 2001; Zhou et al.\ 2015; Pan et al.\ 2016).

The search for QPOs  in light curves of AGN is very important: their presence can provide strong support 
for the common nature of the accretion process onto BHs ranging from a few solar masses up to the SMBHs 
present in quasars (e.g.,\ Abramowicz \& Klu{\'z}niak 2001; Remillard \& McClintock 2006; Zhou et al.\ 2015). 
Plausible  models that might explain  QPOs in AGN in different wavebands and on diverse timescales have 
been put forward (e.g.,\ Gierli{\'n}ski et al.\ 2008; Gupta et al.\ 2009; Lachowicz et al.\ 2009; 
Pan et al.\ 2016; Bhatta 2017, and references therein). However, only once we have at least a handful of good 
cases of QPOs in different subclasses of AGN and on diverse timescales will it be possible for us to gain a 
thorough understanding of this phenomenon.

\begin{figure}
\centering
\includegraphics[scale=0.48]{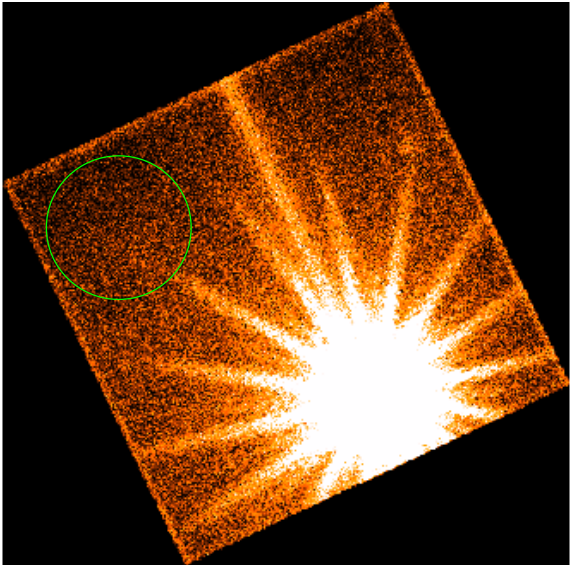} \\
\caption{XMM-Newton EPIC/pn image of the NLSy1 MCG--06--30--15 and the selected background region, denoted by a circle.}
\end{figure}

\begin{figure*}
\centering
\includegraphics[scale=0.90]{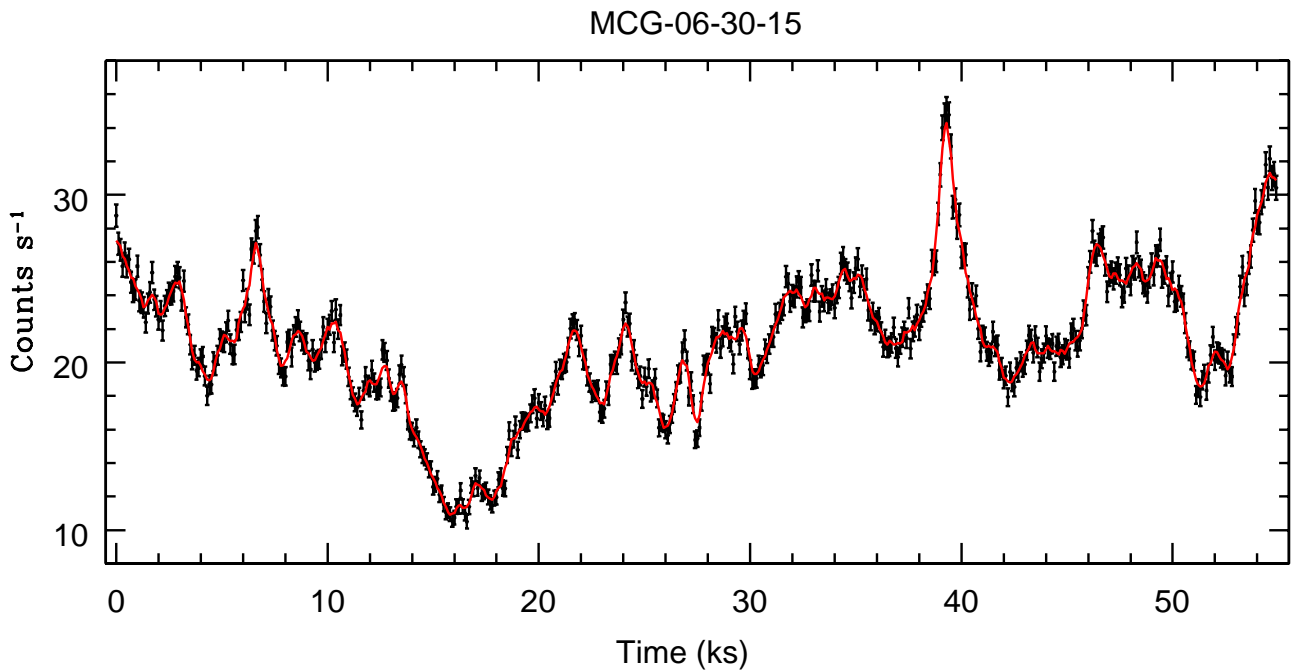}
\vspace*{-3.5in}
\caption{Light curve of NLS1 MCG--06--30--15 in 0.3 -- 10 keV observed with {\it XMM-Newton} EPIC/pn, with a
running average over 5 points given by the  continuous red curve.} 
\end{figure*}

In Section 2 of this Letter we briefly  describe the X-ray data and how we reduced it. In Section 3 we present 
a description of the  QPO search techniques we employed and the results of those analyses. A discussion and 
our conclusions are given in Section 4.

\section{Data and reduction}

We examined the eight archival {\it XMM-Newton} EPIC/pn observations of the NLSy1 galaxy MCG--06--30--15 that had 
high quality data and extended for at least 30 ks.  In total, these exposures amounted to 747 ks.  The observation 
of the greatest interest was taken on 2000 July 11-12 (Orbit 108, Observation ID 0111570201) and lasted 55 ks. 
Wilms et al.\ (2001) analyzed the spectrum of this particular observation combined with others taken on the same 
days. These authors concluded that the reflection continuum was extremely broad, supporting the claim that 
the SMBH is rotating rapidly.

For data reduction, we used the {\it XMM-Newton} Science Analysis Software (SAS), version 15.0.0. We limited our
analysis to EPIC/pn data because it is free from both soft-proton flaring events and pile-up effects. The light curve in
the energy range 0.3 -- 10.0 keV was extracted using a 40 arcsec radius selection region and corrected for background
flux by selecting a 45 arcsec radius region as far as possible from the source on the same CCD chip. Finally, the
$\sim$ 55 ks data were evenly sampled in time bins of 100 s.  The mean count rate and rms variability were found to
be 21.2 ct s$^{-1}$ and 19.52$\pm$0.12\%, respectively.

\section{Light curve analysis and results}

The XMM-Newton EPIC/pn image of NLSy1 MCG--06--30--15 and a circle showing the 45 arc sec radius background region are 
shown in Fig.\ 1. The $\sim$ 55 ks continuous light curve of NLSy1 MCG--06--30--15 in 0.3 -- 10 keV is plotted 
in Fig.\ 2. By visual examination it appears that  the X-ray emission during that period may include a quasi-periodic 
component. To examine and quantify this possibility we analyzed the data using 
recently employing the extensively used Lomb-Scargle periodogram (LSP), and weighted wavelet Z-transform (WWZ) analysis 
techniques. In the following subsections we briefly describe these techniques and the putative QPO periods they yield.

\subsection{Lomb-Scargle periodogram}

The LSP is a powerful technique used to analyze the periodicities in irregular 
time series (Lomb 1976; Scargle 1982), which employs $\chi^2$ statistics to fit sine waves throughout the 
data train. It  reduces the effects of irregular sampling and indicates any periodicities or quasi-periods 
that may be present in the data and also provides their statistical significance. The significance of a 
certain signal is determined by the probability, $p$, of the null hypothesis, which can be calculated using 
the LSP (Hong et al. 2018; Zhang et al. 2017a, 2017b; Zhang et al. 2017). 
The LSP at a particular frequency $\omega_i$, given $N$ observations, is defined as

\begin{displaymath}
P(\omega_i) = {\frac{1}{2 \sigma^2}} \Bigl\{ \frac {(\sum^N_{j=1} [a(t_j)-\bar{a}] \cos[\omega_i(t_j-\tau)])^2}{\sum^N_{j=1} \cos^2[\omega_j(t_j-\tau)]} 
\end{displaymath}
\begin{equation}
+ \frac {(\sum^N_{j=1} [a(t_j)-\bar{a}] \sin[\omega_i(t_j-\tau)])^2} {\sum^N_{j=1} \sin^2[\omega_i(t_j-\tau)]} \Bigr\}
\end{equation}
In this work $i = 1 \dots M$, where $\tau$ is calculated as

\begin{equation}
\tan(2 \omega_i \tau) = \frac{\sum_{j=1}^N \sin(2 \omega_i t_j)}{\sum_{j=1}^N \cos(2 \omega_i t_j)}
\end{equation}
and $M$ is the number of independent frequencies. The significance of features of a periodogram can be assessed 
by testing the null hypothesis of no period being present, parameterized by its $p$-value known as false alarm 
probability (FAP). The $p$-value for the probability that the measured peak is smaller than $y$ is given by
\begin{equation}
p (y)\approx N{\rm exp}(-y),
\end{equation}
where N is the number of data points. The smaller the $p$-value, the higher the significance of any peak.

In the top panel of Fig.\ 3, a periodogram power spectral density for MCG--6--30--15 is plotted against 
frequency using the LSP analysis and the horizontal lines represent $p=0.01$ 
significance levels. The line with $p=\alpha$ corresponds to a $(1-\alpha)\times~100\%$ confidence level. 
A peak of period $3600\pm229$ s is found to have greater than a $99\%$ global significance, signaling a 
possible QPO. Another peak of $7300\pm425$ s signals the possible harmonic of the periodic oscillation 
detected by this method.

In general, the light curves of AGNs are mainly comprised of red noise that arises from stochastic processes in the
 accretion disk or jet and is frequency dependent (Fan et al. 2014; Xiong et al. 2017; Hong et al. 2018). 
 In order to assess the red noise, we employed the REDFIT program which calculates the red noise present in the 
 data by fitting it to the first order auto-regressive (AR1) process (Schulz \& Mudelsee (2002)). The bottom panel 
of Fig.\ 3 plots the REDFIT result, i.e., the spectra, theoretical red noise spectrum, and the $90\%$ significance level. 
The peak at ($2.7\pm0.1)\times 10^{-4}$ indicates the possible periodicity of 3670 s, which is also shown by LSP method. 

\begin{figure}[t]
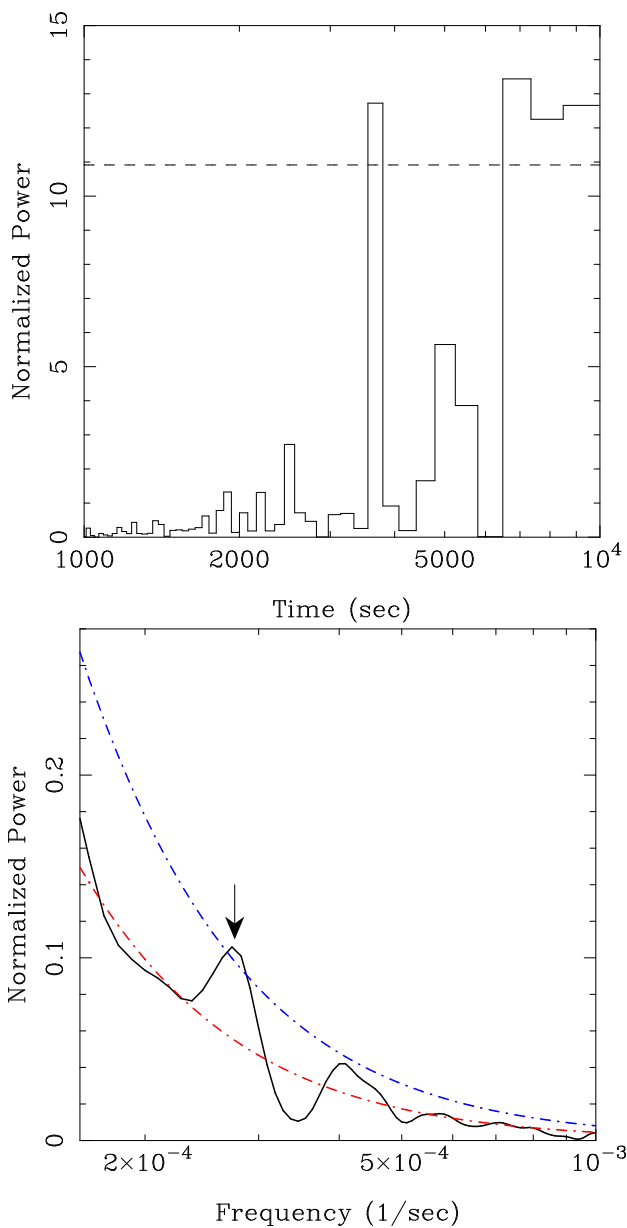

\centering
\includegraphics[scale=0.53]{fig3a.eps} \\
\includegraphics[scale=0.53]{fig3b.eps}
\caption{ {\bf Top panel}: LSP of the light curve in Fig.\ 1. The dashed line represents $p=0.01$. 
{\bf Bottom panel}: Results of REDFIT method. The black line represents the bias corrected spectra, 
the red line indicates the theoretical (AR1) red noise spectrum, and the blue line shows the 90$\%$ $\chi^{2}$ 
significance level.}
\end{figure}

\subsection{Wavelet analysis}

Sometimes  (quasi-) periodic oscillations in a signal do not persist throughout the full dataset.  This is possible 
if there are  QPOs that  evolve with time in their amplitudes and/or frequencies.  The wavelet technique 
measures the non-stationarity of the time series by decomposing it into frequency/time space simultaneously 
(Torrence \& Compo 1998). The WWZ is commonly used for quantifying 
periodicities in the time series and  uses the $z$-statistics of Foster (1996). We employed WWZ software to 
calculate the WWZ power as a function of frequency and time (e.g., King et al. 2013; Bhatta et al. 2016; 
Bhatta 2017; Zhang et al. 2017a, 2017b, 2018; and references therein). 
The color scaled density plot of WWZ is presented in Fig. 4. The figure features concentrations of WWZ 
power around periods of 3665, 5125, and 8736 s with errors of $\sim$180 s for the first two and $\sim 350$ s for the last, 
indicating  possible QPOs in the signal. 
The QPO with period around 3665 seconds gradually develops with moderate strength and is most clearly persistent though 
the second half of the observation. The same is true for a possible QPO around 5125 s but it is stronger at 
the beginning of the observation than the previous QPO. Another possible QPO, which has a period centered at 8736 s, 
on the other hand, is strongest and more persistent throughout  the observation. The right-hand panel 
of Fig.\ 4 plots the time averaged WWZ against period and shows peaks at those three values.

\section{Discussion and conclusions}

The quasi-periods obtained from the LSP and WWZ analyses, which search for the presence of  sinusoidal components, 
agree to within the resolutions of these techniques.  Together, they  indicate that during this particular 
observation the NLSy1 MCG--06--30--15 exhibited a QPO at a frequency of $ \simeq 2.73 \times 10^{-4}$ Hz, or 
at a central period of $\simeq 3670$~s. The LSP approach is the most common frequency-based
technique and can provide good estimates of the strength of a signal if the underlying power spectrum 
can be sensibly modeled, but it does not take 
into account any possible evolution of a periodic or quasi-periodic signal
with time.  The time-frequency approach using wavelets has the advantage of quantifying the persistence 
of such signals in the data.

Because we found no indication of a QPO in the other seven  extensive 
light curves for this source that we examined, one may consider that the formal statistical confidence in 
any long-lived QPO is reduced to $\sim$ 90\%. However, we 
believe that it is more appropriate to interpret this result as being consistent with the general 
finding  that detectable QPOs from AGN are not long-lived, but rather are transient in nature 
(Gierli{\'n}ski et al.\ 2008; Pan et al.\ 2016), and thus this object may have a QPO duty cycle of $\sim$12\%.  
We note that our signal appears to be present throughout the full observation of MCG--06--30--15; 
this is in contrast to the QPOs detected in the other two NLSy1s,  RE J1034$+$396 
(Gierli{\'n}ski et al.\ 2008; Czerny et al.\ 2010) and 1H 0707$-$495 (Pan et al.\ 2016), in which the QPOs were only 
significant in portions of the data trains.

Recently the SMBH in MCG--06--30--15 has had its mass estimated in two independent direct ways  
using the reverberation mapping technique for the H$\beta$ line.  Hu et al.\ (2016) found a 
preferred value of $3.26^{+1.59}_{-1.40} \times 10^6 M_{\odot}$ but Bentz et al.\ (2016) obtained 
$1.6 \pm 0.4 \times 10^6 M_{\odot}$ using different reverberation campaigns and somewhat different analysis 
techniques.  The time lags these two groups  measure at multiple epochs are reasonably consistent, at 6.38 d 
and 5.33 d, respectively, as are their values of the H$\beta$ FWHM, but Hu et al.\ (2016) used the FWHM and 
Bentz et al.\ (2016) used the dispersion, $\sigma$.   These choices, along with different choices for the 
value of the scaling factor that accounts for the kinematics and geometry of the broad-line region gas, 
produce the differences in their best mass values. These values, nevertheless, are still consistent within 
the errors. These papers estimated Eddington ratios of $\sim 0.12$ (Hu et al.\ 2016) and $\sim 0.04$ 
(Bentz et al.\ 2016), which are substantial, but lower than that for most NLSy1s (Hu et al.\ 2016). Most 
of the peculiar properties of NLSy1s can be understood if they are Seyferts with relatively low SMBH 
masses that are accreting at relatively high rates.

High-frequency QPOs (HFQPOs) for BH X-ray binaries are in the range 40--450 Hz and because their 
frequencies are constant despite large changes in luminosity it has been thought that they are related 
to the innermost portions of the accretion disks and hence may provide measures of the masses and spins 
of their BHs (e.g.,\ Abramowicz \& Klu{\'z}niak 2001; Abramowicz et al.\ 2004; Remillard \& McClintock 2006).   
A plot of HFQPO frequencies against  BH mass produces the expected tight inverse relationship for 
stellar mass BHs and this has recently been extended from Galactic BH QPOs through possible intermediate 
mass BHs up to SMBHs (Zhou et al.\ 2015, Pan et al.\ 2016, and references therein).  The frequency of this
probable QPO in MCG--06--30--15 is nearly identical to those found in RE J1034$+$396 and 1H 0707$-$495 
at $\simeq 2.7 \times 10^{-4}$Hz. The best mass estimates of this QPO, 
at $3.26^{+1.59}_{-1.40} \times 10^6 M_{\odot}$ or $1.6 \pm 0.4 \times 10^6 M_{\odot}$,  are also very 
close to the others,  which are $4^{+3}_ {-1.5} \times 10^6 M_{\odot}$ for RE J1034$+$396 (Zhou et al.\ 2010)
and $5.2(\pm0.5{\rm dex}) \times 10^6 M_{\odot}$ for 1H 0707$-$495 (Pan et al.\ 2016).  Hence, our new 
measurement appears to help confirm the inverse linear dependence of QPO frequency and BH mass and would 
lie right next to the other AGN points on Fig.\ 4 of Pan et al.\ (2016).

\begin{figure*}
\centering
\includegraphics[angle=90, scale=0.63]{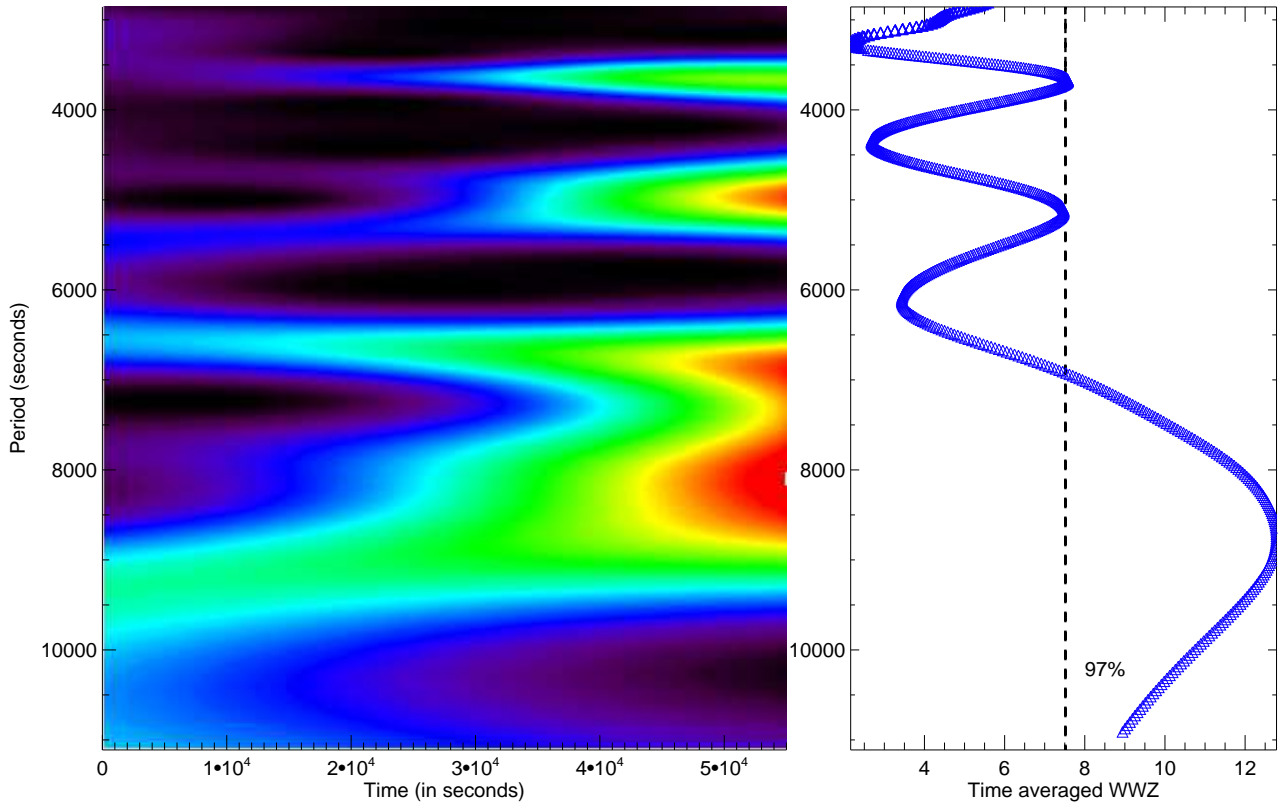}
\caption  {Weighted wavelet z-transform of the light curve presented in Fig. 1. The left panel shows the 
distribution of color-scaled WWZ power (with red most intense) in the time-period plane, and the right panel 
shows the time-averaged WWZ power (blue curve) as a function of period. The dotted black curve represents 
$97\%$ global significance.}
\end{figure*}

This tight bunching of frequencies (and SMBH masses) for AGN QPOs may seem surprising but selection 
effects probably strongly favor QPO frequencies and SMBH masses relation. The higher the SMBH mass the 
longer the expected possible QPO period and the longer any observation would need to be to have a chance 
of detecting a QPO. Of course only Type 1 AGN afford us the opportunity of seeing the region close to the 
BH from which QPOs emerge. Those BHs are likely to be several times more massive in normal Sy1s than in the 
NLSy1s in which QPOs have been seen.  Most QSOs have much more massive SMBHs, and while they are also much 
more luminous, they are also much further away. Therefore count rates are no greater and the required 
continuous observation times needed to perform good QPO searches on much longer timescales are essentially 
impossible to obtain.  

The presence of 3:2 ratios for the frequencies of many HFQPOs in X-ray binaries indicates that 
the physical mechanism producing these HFQPOs involves a resonance phenomenon of some sort 
(e.g.,\ Abramowicz \& Klu{\'z}niak 2001) and the scaled similarity between those stellar mass systems 
and these NLSy1 AGNs supports the idea that resonances are important for AGNs as well.  Although no 
pairs of QPOs at that 3:2 ratio (or any other, for that matter) have yet been detected for any AGN 
and the discovery of such would be extraordinarily important, the resonance hypothesis certainly  is  
reasonable.  An estimated Eddington ratio of 0.12 is significantly higher than that of the great 
majority of AGN. This large Eddington ratio and the likelihood that this is another HFQPO
supports the claim by Pan et al.\ (2016) that HFQPOs are associated with BHs accreting at very high 
rates as is the case for BH X-ray binaries.  If  a high accretion rate is indeed a prerequisite for 
engendering a HFQPO and the few detected in AGNs are indeed of the same type as in X-ray binaries, 
then many otherwise viable models  (e.g., review by Belloni \& Stella 2014) 
are disfavored (Pan et al.\ 2016).  A wavelet analysis of the apparent variations of the QPO 
frequency for the NLS1 RE J1034$+$396 by Czerny et al.\ (2010) indicated that an increase in QPO 
frequency was accompanied by an increase  in X-ray flux.  The wavelet approach to the current 
observation of  MCG--06--30--15 does not provide additional evidence for such a trend, as the 
QPO frequency around 3670 s appears to be very stable. According to Pan et al.\ (2016), only 
models invoking p-modes trapped in the innermost part of an accretion disk (e.g., Li et al.\ 2003, 
Hor{\'a}k \& Lai 2013, and references therein) or those involving magnetized disks subject 
to accretion-ejection instabilities (Tagger \& Pellat 1999) naturally tie HFQPOs to high accretion 
rate situations. 

We thank the anonymous referee for constructive comments and suggestions. ACG thanks Prof.\ D.\ Banerjee 
for providing a computer code for wavelet analysis. ACG is partially supported by the CAS President's 
International Fellowship Initiative (PIFI), Grant No. 2016VMB073.  AT acknowledges support from 
the China Scholarship Council (CSC), Grant No.\ 2016GXZR89. PJW is grateful for hospitality at SHAO 
while this paper was written. MFG is supported by the National Science Foundation of China (Grant 
Nos.\ 11473054 and U1531245) and by the Science and Technology Commission of Shanghai Municipality 
(Grant No.\ 14ZR1447100). CB was supported by the National Natural Science Foundation of China 
(Grant No.\ U1531117), Fudan University (Grant No.\ IDH1512060), and the Alexander von Humboldt 
Foundation. LCH was supported by the National Key R\&D Program of China (2016YFA0400702) and the 
National Science Foundation of China (11473002, 11721303). 

This research is based on observations obtained with {\it XMM-Newton}, an ESA science mission
with instruments and contributions directly funded by ESA Member States and NASA. WWZ software is 
available at URL:  https://www.aavso.org/software-directory.

\end{document}